\documentclass[a4paper,10pt]{article}
\usepackage{amssymb,latexsym,amsmath,amscd}

\numberwithin{equation}{section}
\numberwithin{figure}{section}

\usepackage{hyperref}
\usepackage{graphicx}
\usepackage[usenames,dvipsnames]{pstricks}
\usepackage{epsfig}

\def\ap{\alpha'}
\def\set1#1{\{\, #1 \,\}}
\def\zz{\mathbb Z}

\def\rr{\mathbb R}

\title{{\bf $A_\infty$-structure in string theory and \\ the Yang-Mills equation}}
\date{}
\author{{\bf Dmitry Grigoryev}\footnote{Steklov Mathematical Institute, Moscow, Russia. e-mail dmitriy57@gmail.com}\; and
	  {\bf Pavel Khromov}\footnote{Steklov Mathematical Institute, Moscow, Russia. e-mail pasha.khromov@gmail.com}}
\begin{document}
\maketitle
\begin{abstract}
We consider local operators of CFT inserted at the boundary of the worldsheet
and an infinite set of maps that act on a space of the local operators.
These maps have natural CFT interpretation and form $A_\infty$-algebra.
In terms of these operators we define the homotopical Maurer-Cartan equation,
find its symmetries and explore its properties.
Further we recover the Yang-Mills equation from the homotopical Maurer-Cartan equation,
identify the first $\ap$-correction to it
and propose method for calculation all corrections.
\end{abstract}

\section{Introduction}
String theory on a nontrivial background is described in terms of essentially nonlinear sigma-models.
For a string to consistently propagate in the background
the corresponding sigma-model must be conformal invariant.
Condition of conformal invariance turns out to be an equations of motion to a fields on the target space of the theory.
Thus there is a well-known statement \cite{FT-1,Polyakov}:
equation of motion for fields on a target space of a sigma-model
is a beta function vanishing conditions for the corresponding
two dimensional sigma-model.
In the 80th using sigma model approach
classical equations of motion were reproduced and
$\ap$-corrections the equations were calculated \cite{FT-2}-\cite{Tseytlin}.

In this paper we check the Losev's conjecture.
It says that particular representations of the homotopical Maurer-Cartan (hMC) equation
reproduce classical equations of motion (Yang-Mills or Einstein, for example) together with the string corrections.
By the hMC equation we mean a generalization of common Maurer-Cartan equation to $A_\infty$-algebras.
These algebras were firstly introduced by Stasheff in \cite{Stasheff-1}.
Review on issues concerning $A_\infty$-structure may be found in \cite{A-inf} and references therein.
Higher structures find applications in closed string field theory \cite{Zwiebach},
topological theories \cite{Kontsevich,Tomasiello}
and in study of gauge theories and gravity via the hMC equation
that was suggested in \cite{1-formalism} and developed in \cite{KS-1,KS-2,Zeitlin}.

To check the conjecture we consider
local operators inserted at the boundary of the worldsheet
and a set of maps that acts on the space of the local operators.
The first map is the BRST differential,
the second is a binary map
that is just an operator product expansion
(depending on the splitting between the operators).
It turns out that these two maps may be completed with the
infinite set of maps, also depending on a single parameter, such that they form an $A_\infty$-algebra,
thus depending on the parameter too.

In this setup using a single proposal we
reproduce the Yang-Mills equation from the hMC equation
and identify the first $\ap$-correction that is linear in $\ap$.
This approach of course can be applied in a similar way to a calculation of $\ap$-corrections of degree more that one,
but in this paper we have restricted ourselves to a discussion of the main idea and the simplest example.

The paper has the following structure:
in section~\ref{section:hMC} definition of maps that form $A_\infty$-algebra is given.
Then we study properties of the hMC equation written in terms of this maps.
In section~\ref{section:YM} using a single proposal we reproduce the Yang-Mills equation and calculate the first $\ap$-correction to it from the hMC equation.
Important but quite technical issues may be found in the Appendix.
Some speculations concerning suggested approach are collected in the Conclusion.

\section{The homotopical Maurer-Cartan equation}\label{section:hMC}
To discuss the properties of the equation we first need to define
maps in terms of which the equation will be written.

All our discussion will be associated with the following CFT
\begin{equation}
  S=\frac{1}{2\pi\ap}\int_\Sigma d^2z\; \eta_{\mu\nu}\partial X^\mu\partial X^\nu + \int_\Sigma d^2z\; b\bar\partial c +c.c.
\end{equation}
which governs a maps from the open string worldsheet $\Sigma$ to the Minkowski space of dimension 26 with the standard flat metric on it.
For the fields of the theory we have standard OPEs.

\subsection{The maps}
Let $\varphi$ be a local operator inserted at the boundary of the worldsheet.
In general we will be dealing with Lie algebra valued local operators, so they don't commute.
The open string worldsheet may be brought to the upper half plane (UHP) of the complex plane by a proper conformal mapping.
In this representation the local operators are inserted at the real line.

Let $\Omega^n$ be the space of local operators with ghost number $n$. Then we have the following complex:
\begin{equation}\label{diagr}
\begin{CD}
\ldots @>Q>>\Omega^0 @>Q>> \Omega^1 @>Q>> \Omega^2 @>Q>> \ldots \\
\end{CD}
\end{equation}
with a differential which is the BRST operator.

We consider a space of all operators $\mathcal H=\oplus_{n\in\zz}\Omega^n$ and
introduce $A_\infty$-algebra of maps $M_1$, $M_2$,\dots that act on this space:
\begin{equation}
  M_k: \mathcal H^{\otimes k}\rightarrow\mathcal H.
\end{equation}
Here we follow the following route to define maps.
First, we introduce geometric definition of maps $M_k$ that depend on a single parameter.
In terms of these maps we can define the hMC equation.
As long as the definition of the maps $M_k$ involves notion of complicated Riemann surfaces,
these maps are inconvenient for doing practical calculations.
So we map these maps to a familiar region, namely the UHP, and define $m_k=f\circ M_k$, where $f$ is a proper conformal mapping to the UHP,
which is different for different maps $M_k$.
Next we provide an evidence that in some cases -- and the calculation we do to reproduce the Yang-Mills belongs to this case --
we even don't need to know the exact mapping,
but all we need to know is a little piece of information about the mapping (that lead to appearance of splittings)
and some extra assumption (that lead to hierarchy).

\begin{figure}[h!]
\begin{center}
\begin{picture}(0,0)%
\includegraphics{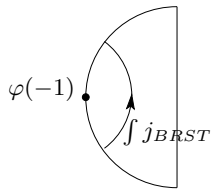}%
\end{picture}%
\setlength{\unitlength}{1243sp}%
\begingroup\makeatletter\ifx\SetFigFont\undefined%
\gdef\SetFigFont#1#2#3#4#5{%
  \reset@font\fontsize{#1}{#2pt}%
  \fontfamily{#3}\fontseries{#4}\fontshape{#5}%
  \selectfont}%
\fi\endgroup%
\begin{picture}(3357,3624)(-1094,-6823)
\put(1171,-5821){\makebox(0,0)[lb]{\smash{{\SetFigFont{9}{10.8}{\rmdefault}{\mddefault}{\updefault}{\color[rgb]{0,0,0}$\int j_{BRST}$}%
}}}}
\put(-1079,-4966){\makebox(0,0)[lb]{\smash{{\SetFigFont{9}{10.8}{\rmdefault}{\mddefault}{\updefault}{\color[rgb]{0,0,0}$\varphi(-1)$}%
}}}}
\end{picture}%
\caption{To the definition of $M_1$.}
\label{fig:m1}
\end{center}
\end{figure}

We define the first map $M_1$ in the following way:
\begin{align}
  M_1(\varphi)=Q(\varphi),
\end{align}
where a local operator $\varphi$ is inserted at the point $-1$ on a half disc (see fig. \ref{fig:m1}).

Then we define
\begin{align}
  m_1=f\circ M_1,	
\end{align}
which is an action of the BRST operator on the image of $\varphi$ on the UHP.
For a primary local operator we can write
\begin{align}
  f\circ M_1(\varphi(-1))=CF\cdot M_1(\varphi(f(-1))),
\end{align}
where $CF$ denotes a conformal factor that arises due to transformational properties of operators we are acting on.

\begin{figure}[ht!]
  \centering
  \parbox{4in}{
    \centering
    \begin{picture}(0,0)%
\includegraphics{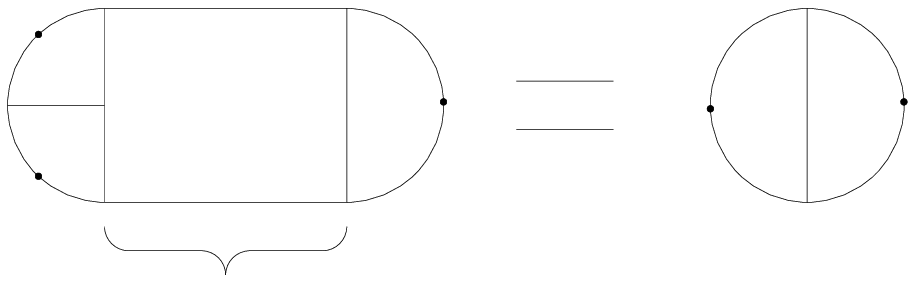}%
\end{picture}%
\setlength{\unitlength}{1243sp}%
\begingroup\makeatletter\ifx\SetFigFont\undefined%
\gdef\SetFigFont#1#2#3#4#5{%
  \reset@font\fontsize{#1}{#2pt}%
  \fontfamily{#3}\fontseries{#4}\fontshape{#5}%
  \selectfont}%
\fi\endgroup%
\begin{picture}(13890,4753)(706,-4607)
\put(721,-241){\makebox(0,0)[lb]{\smash{{\SetFigFont{9}{10.8}{\rmdefault}{\mddefault}{\updefault}{\color[rgb]{0,0,0}$\varphi_1$}%
}}}}
\put(811,-3031){\makebox(0,0)[lb]{\smash{{\SetFigFont{9}{10.8}{\rmdefault}{\mddefault}{\updefault}{\color[rgb]{0,0,0}$\varphi_2$}%
}}}}
\put(2341,-1591){\makebox(0,0)[lb]{\smash{{\SetFigFont{9}{10.8}{\rmdefault}{\mddefault}{\updefault}{\color[rgb]{0,0,0}$S$}%
}}}}
\put(3871,-4471){\makebox(0,0)[lb]{\smash{{\SetFigFont{9}{10.8}{\rmdefault}{\mddefault}{\updefault}{\color[rgb]{0,0,0}$\tau$}%
}}}}
\put(7561,-1591){\makebox(0,0)[lb]{\smash{{\SetFigFont{9}{10.8}{\rmdefault}{\mddefault}{\updefault}{\color[rgb]{0,0,0}$\chi$}%
}}}}
\put(10441,-1591){\makebox(0,0)[lb]{\smash{{\SetFigFont{9}{10.8}{\rmdefault}{\mddefault}{\updefault}{\color[rgb]{0,0,0}$\varphi_{out}$}%
}}}}
\put(14581,-1591){\makebox(0,0)[lb]{\smash{{\SetFigFont{9}{10.8}{\rmdefault}{\mddefault}{\updefault}{\color[rgb]{0,0,0}$\chi$}%
}}}}
\end{picture}%
    \caption{To the definition of $M_2$. Left correlator is calculated on a surface glued from three unit half-discs with
      local operator insertions and a strip of width $2$ and length $\tau$. There is a conical singularity
      at the point $S$ with total angle $3\pi$ (each individual angle is $\pi$). Right correlator is calculated on a unit disc.}
    \label{fig:m2}}
  \parbox{1.7in}{
    \centering
    \begin{picture}(0,0)%
\includegraphics{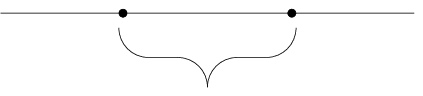}%
\end{picture}%
\setlength{\unitlength}{1243sp}%
\begingroup\makeatletter\ifx\SetFigFont\undefined%
\gdef\SetFigFont#1#2#3#4#5{%
  \reset@font\fontsize{#1}{#2pt}%
  \fontfamily{#3}\fontseries{#4}\fontshape{#5}%
  \selectfont}%
\fi\endgroup%
\begin{picture}(6324,2358)(889,-2695)
\put(3871,-2536){\makebox(0,0)[lb]{\smash{{\SetFigFont{9}{10.8}{\rmdefault}{\mddefault}{\updefault}{\color[rgb]{0,0,0}$\epsilon$}%
}}}}
\put(4951,-736){\makebox(0,0)[lb]{\smash{{\SetFigFont{9}{10.8}{\rmdefault}{\mddefault}{\updefault}{\color[rgb]{0,0,0}$\varphi_2(y)$}%
}}}}
\put(2476,-736){\makebox(0,0)[lb]{\smash{{\SetFigFont{9}{10.8}{\rmdefault}{\mddefault}{\updefault}{\color[rgb]{0,0,0}$\varphi_1(x)$}%
}}}}
\end{picture}%
    \caption{To the definition of $m_2$.}
    \label{fig:m2-uhp}}
\end{figure}

We define the second map $M_2$ in the following way:
\begin{equation}\label{m2-mapping}
  M_2(\varphi_1(w_1),\varphi_2(w_2))=e^{-\tau L_0} \varphi_1(w_1)\varphi_2(w_2),
\end{equation}
where $L_0$ is a 0-th mode of stress-energy tensor of the CFT and
the exponent acts on all operators that stand to the right.

As long as
$e^{-\tau L_0}$
represents an insertion of a strip with a length $\tau$
we may think about result of application $\varphi_{out}=M_2(\varphi_1,\varphi_2)$ of $M_2$ to operators $\varphi_1$ and $\varphi_2$
as a correlator calculated on a Riemann surface glued from two half-discs and the $\tau$-strip (see fig. \ref{fig:m2})
where equality holds for any operator $\chi$.

Let us note that at $\tau=0$ the map $M_2$ coincides with the Witten's product \cite{Witten-sft} of string states corresponding
to local operators $\varphi_1$ and $\varphi_2$  in the open string field theory,
which is non-commutative and associative.
We study all maps with $\tau\ne 0$ so $M_2$ is non-commutative and non-associative.

We define
\begin{align}
  m_2 = f\circ M_2
\end{align}
with $f$ being a conformal mapping to the UHP.
For a primary local operators we can write
\begin{align}\label{m2-uhp-mapping}
  f\circ M_2\bigl(\varphi_1(z_1),\varphi_2(z_2)\bigr)= CF \cdot m_2\bigl(\varphi_1(f(z_1)),\varphi_2(f(z_2))\bigr),
\end{align}
where again $CF$ denotes a conformal factor that arises due to transformational properties of operators we are acting on.
But we actually aren't interested in the images of $z_1$ and $z_2$ under the mapping $f$.
So let us write $f(z_1)=x$ and $f(z_2)=x+\epsilon$.
We call parameter $\epsilon$ splitting.
Presence of $\tau$-strip in \eqref{m2-mapping} results in behavior $\epsilon\sim e^{-\tau}\to0$
of the splitting between the local operators in \eqref{m2-uhp-mapping} in a limit $\tau\to\infty$ (see fig. \ref{fig:m2-uhp}).
Thus we can write
\footnote{In general one should keep regular terms in splittings
because they might cancel out divergences arising from contractions.
But one can convince himself that this happens at a higher (more than 3) orders in the coupling $t$ (see section \ref{section:YM}), 
so this situation is crucial only when calculating $\ap$-corrections.}
\begin{align}
  m_2^\epsilon(\varphi_1,\varphi_2)(x) =  CF \cdot \Bigl[\varphi_1(x)\varphi_2(x+\epsilon)\Bigr]_x
\end{align}
with the notation $[\mathcal O(x_1,...,x_k)]_{x_1}$ meaning that we expand the expression $\mathcal O(x_1,...,x_k)$ at the point $x_1$.
So applying this map to operators $\varphi_1$ and $\varphi_2$ we simply perform OPE of $\varphi_1(x)\varphi_2(x+\epsilon)$ at the point $x$.

\begin{figure}[!h]
  \centering
  \parbox{2.8in}{
    \centering
    \begin{picture}(0,0)%
\includegraphics{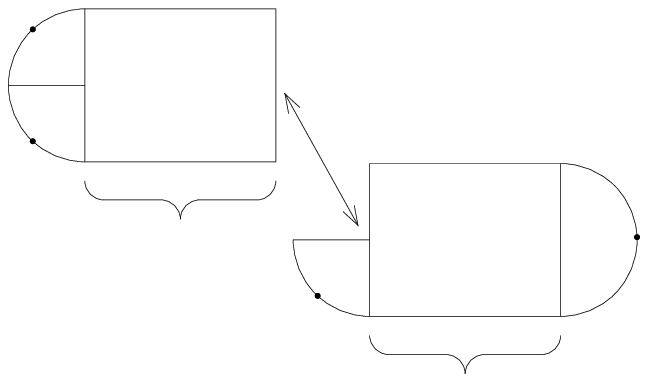}%
\end{picture}%
\setlength{\unitlength}{1243sp}%
\begingroup\makeatletter\ifx\SetFigFont\undefined%
\gdef\SetFigFont#1#2#3#4#5{%
  \reset@font\fontsize{#1}{#2pt}%
  \fontfamily{#3}\fontseries{#4}\fontshape{#5}%
  \selectfont}%
\fi\endgroup%
\begin{picture}(9840,6373)(616,-6047)
\put(10441,-3391){\makebox(0,0)[lb]{\smash{{\SetFigFont{9}{10.8}{\rmdefault}{\mddefault}{\updefault}{\color[rgb]{0,0,0}$\chi$}%
}}}}
\put(7471,-5911){\makebox(0,0)[lb]{\smash{{\SetFigFont{9}{10.8}{\rmdefault}{\mddefault}{\updefault}{\color[rgb]{0,0,0}$\tau$}%
}}}}
\put(5041,-4651){\makebox(0,0)[lb]{\smash{{\SetFigFont{9}{10.8}{\rmdefault}{\mddefault}{\updefault}{\color[rgb]{0,0,0}$\varphi_3$}%
}}}}
\put(721,-2221){\makebox(0,0)[lb]{\smash{{\SetFigFont{9}{10.8}{\rmdefault}{\mddefault}{\updefault}{\color[rgb]{0,0,0}$\varphi_2$}%
}}}}
\put(3151,-3481){\makebox(0,0)[lb]{\smash{{\SetFigFont{9}{10.8}{\rmdefault}{\mddefault}{\updefault}{\color[rgb]{0,0,0}$\tau$}%
}}}}
\put(631,-61){\makebox(0,0)[lb]{\smash{{\SetFigFont{9}{10.8}{\rmdefault}{\mddefault}{\updefault}{\color[rgb]{0,0,0}$\varphi_1$}%
}}}}
\end{picture}%
    \caption{Gluing together two surfaces to evaluate one way of a composition $M_2\circ M_2$.}
    \label{fig:m2-m2-glue}}
  \parbox{2.8in}{
    \centering
    \begin{picture}(0,0)%
\includegraphics{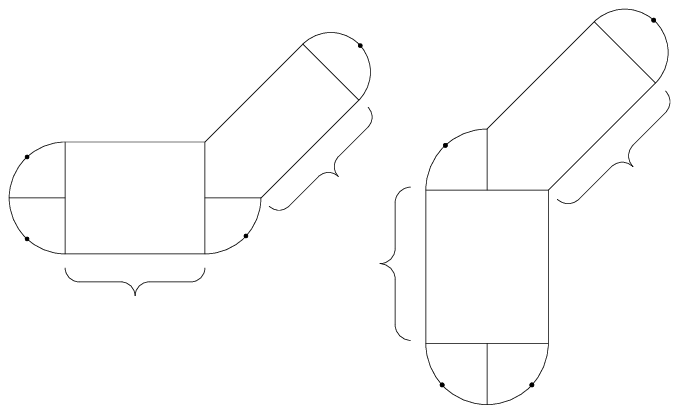}%
\end{picture}%
\setlength{\unitlength}{1243sp}%
\begingroup\makeatletter\ifx\SetFigFont\undefined%
\gdef\SetFigFont#1#2#3#4#5{%
  \reset@font\fontsize{#1}{#2pt}%
  \fontfamily{#3}\fontseries{#4}\fontshape{#5}%
  \selectfont}%
\fi\endgroup%
\begin{picture}(10249,6568)(1201,-7370)
\put(6571,-5101){\makebox(0,0)[lb]{\smash{{\SetFigFont{9}{10.8}{\rmdefault}{\mddefault}{\updefault}{\color[rgb]{0,0,0}$\tau$}%
}}}}
\put(7471,-3121){\makebox(0,0)[lb]{\smash{{\SetFigFont{9}{10.8}{\rmdefault}{\mddefault}{\updefault}{\color[rgb]{0,0,0}$\varphi_1$}%
}}}}
\put(9451,-7171){\makebox(0,0)[lb]{\smash{{\SetFigFont{9}{10.8}{\rmdefault}{\mddefault}{\updefault}{\color[rgb]{0,0,0}$\varphi_3$}%
}}}}
\put(8731,-3211){\makebox(0,0)[lb]{\smash{{\SetFigFont{9}{10.8}{\rmdefault}{\mddefault}{\updefault}{\color[rgb]{0,0,0}$S_3$}%
}}}}
\put(11341,-1141){\makebox(0,0)[lb]{\smash{{\SetFigFont{9}{10.8}{\rmdefault}{\mddefault}{\updefault}{\color[rgb]{0,0,0}$\chi$}%
}}}}
\put(10801,-3841){\makebox(0,0)[lb]{\smash{{\SetFigFont{9}{10.8}{\rmdefault}{\mddefault}{\updefault}{\color[rgb]{0,0,0}$\tau$}%
}}}}
\put(7651,-7216){\makebox(0,0)[lb]{\smash{{\SetFigFont{9}{10.8}{\rmdefault}{\mddefault}{\updefault}{\color[rgb]{0,0,0}$\varphi_2$}%
}}}}
\put(9136,-3751){\makebox(0,0)[lb]{\smash{{\SetFigFont{9}{10.8}{\rmdefault}{\mddefault}{\updefault}{\color[rgb]{0,0,0}$S_4$}%
}}}}
\put(8416,-6091){\makebox(0,0)[lb]{\smash{{\SetFigFont{9}{10.8}{\rmdefault}{\mddefault}{\updefault}{\color[rgb]{0,0,0}$S_1$}%
}}}}
\put(8416,-4291){\makebox(0,0)[lb]{\smash{{\SetFigFont{9}{10.8}{\rmdefault}{\mddefault}{\updefault}{\color[rgb]{0,0,0}$S_2$}%
}}}}
\put(1261,-4966){\makebox(0,0)[lb]{\smash{{\SetFigFont{9}{10.8}{\rmdefault}{\mddefault}{\updefault}{\color[rgb]{0,0,0}$\varphi_2$}%
}}}}
\put(4411,-3346){\makebox(0,0)[lb]{\smash{{\SetFigFont{9}{10.8}{\rmdefault}{\mddefault}{\updefault}{\color[rgb]{0,0,0}$S_3$}%
}}}}
\put(6886,-1546){\makebox(0,0)[lb]{\smash{{\SetFigFont{9}{10.8}{\rmdefault}{\mddefault}{\updefault}{\color[rgb]{0,0,0}$\chi$}%
}}}}
\put(6346,-3976){\makebox(0,0)[lb]{\smash{{\SetFigFont{9}{10.8}{\rmdefault}{\mddefault}{\updefault}{\color[rgb]{0,0,0}$\tau$}%
}}}}
\put(1216,-3256){\makebox(0,0)[lb]{\smash{{\SetFigFont{9}{10.8}{\rmdefault}{\mddefault}{\updefault}{\color[rgb]{0,0,0}$\varphi_1$}%
}}}}
\put(4996,-4966){\makebox(0,0)[lb]{\smash{{\SetFigFont{9}{10.8}{\rmdefault}{\mddefault}{\updefault}{\color[rgb]{0,0,0}$\varphi_3$}%
}}}}
\put(3061,-5911){\makebox(0,0)[lb]{\smash{{\SetFigFont{9}{10.8}{\rmdefault}{\mddefault}{\updefault}{\color[rgb]{0,0,0}$\tau$}%
}}}}
\put(2296,-4156){\makebox(0,0)[lb]{\smash{{\SetFigFont{9}{10.8}{\rmdefault}{\mddefault}{\updefault}{\color[rgb]{0,0,0}$S_1$}%
}}}}
\put(3781,-4156){\makebox(0,0)[lb]{\smash{{\SetFigFont{9}{10.8}{\rmdefault}{\mddefault}{\updefault}{\color[rgb]{0,0,0}$S_2$}%
}}}}
\put(4771,-3886){\makebox(0,0)[lb]{\smash{{\SetFigFont{9}{10.8}{\rmdefault}{\mddefault}{\updefault}{\color[rgb]{0,0,0}$S_4$}%
}}}}
\end{picture}%
    \caption{Two ways of composing $M_2\circ M_2$. Each angle at points $S_1$ and $S_2$ is $\pi$ (total angle is $3\pi$). Each angle at points $S_3$ and $S_4$ is $\pi/2$.}
    \label{fig:m2-m2}}
\end{figure}

\begin{figure}[h!]
\begin{center}
  \begin{picture}(0,0)%
\includegraphics{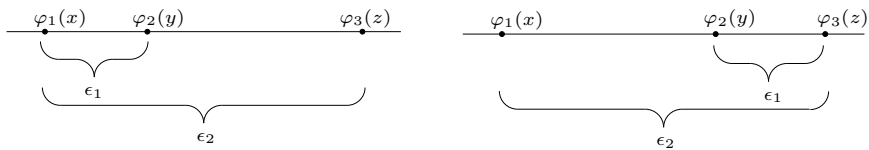}%
\end{picture}%
\setlength{\unitlength}{1243sp}%
\begingroup\makeatletter\ifx\SetFigFont\undefined%
\gdef\SetFigFont#1#2#3#4#5{%
  \reset@font\fontsize{#1}{#2pt}%
  \fontfamily{#3}\fontseries{#4}\fontshape{#5}%
  \selectfont}%
\fi\endgroup%
\begin{picture}(16946,2930)(-11,-2280)
\put(12818,-2139){\makebox(0,0)[lb]{\smash{{\SetFigFont{7}{8.4}{\rmdefault}{\mddefault}{\updefault}{\color[rgb]{0,0,0}$\epsilon_2$}%
}}}}
\put(3749,-2051){\makebox(0,0)[lb]{\smash{{\SetFigFont{7}{8.4}{\rmdefault}{\mddefault}{\updefault}{\color[rgb]{0,0,0}$\epsilon_2$}%
}}}}
\put(2476,299){\makebox(0,0)[lb]{\smash{{\SetFigFont{7}{8.4}{\rmdefault}{\mddefault}{\updefault}{\color[rgb]{0,0,0}$\varphi_2(y)$}%
}}}}
\put(9540,254){\makebox(0,0)[lb]{\smash{{\SetFigFont{7}{8.4}{\rmdefault}{\mddefault}{\updefault}{\color[rgb]{0,0,0}$\varphi_1(x)$}%
}}}}
\put(13752,290){\makebox(0,0)[lb]{\smash{{\SetFigFont{7}{8.4}{\rmdefault}{\mddefault}{\updefault}{\color[rgb]{0,0,0}$\varphi_2(y)$}%
}}}}
\put(15985,290){\makebox(0,0)[lb]{\smash{{\SetFigFont{7}{8.4}{\rmdefault}{\mddefault}{\updefault}{\color[rgb]{0,0,0}$\varphi_3(z)$}%
}}}}
\put(14941,-1231){\makebox(0,0)[lb]{\smash{{\SetFigFont{7}{8.4}{\rmdefault}{\mddefault}{\updefault}{\color[rgb]{0,0,0}$\epsilon_1$}%
}}}}
\put(6571,299){\makebox(0,0)[lb]{\smash{{\SetFigFont{7}{8.4}{\rmdefault}{\mddefault}{\updefault}{\color[rgb]{0,0,0}$\varphi_3(z)$}%
}}}}
\put(531,299){\makebox(0,0)[lb]{\smash{{\SetFigFont{7}{8.4}{\rmdefault}{\mddefault}{\updefault}{\color[rgb]{0,0,0}$\varphi_1(x)$}%
}}}}
\put(1531,-1141){\makebox(0,0)[lb]{\smash{{\SetFigFont{7}{8.4}{\rmdefault}{\mddefault}{\updefault}{\color[rgb]{0,0,0}$\epsilon_1$}%
}}}}
\end{picture}%
  \caption{Compositions $m_2^{\epsilon_2}(m_2^{\epsilon_1}(\varphi_1,\varphi_2),\varphi_3)$ and $m_2^{\epsilon_2}(\varphi_1,m_2^{\epsilon_1}(\varphi_2,\varphi_3))$ correspondingly.}
  \label{fig:m2-m2-uhp}
\end{center}
\end{figure}

Let us find what corresponds to a composition $M_2\circ M_2$ on the UHP.
To do this we
switch to an equivalent definition with Riemann surfaces and
examine properties of $\varphi_{out}=M_2(M_2(\varphi_1,\varphi_2),\varphi_3)$.
The composition can be calculated as follows:
$\varphi_{out}= M_2(\psi,\varphi_3)$, where $\psi=M_2(\varphi_1,\varphi_2)$ is a local operator.
In the correlator $\langle M_2(\psi,\varphi_3),\chi\rangle$ used for defining $\varphi_{out}$ we can replace $\psi$ with
a part of a surface which gives the same correlator (see fig. \ref{fig:m2-m2-glue}).
Since $M_2$ is non-associative there are two ways of composing $M_2\circ M_2$, namely $M_2(M_2(\varphi_1,\varphi_2),\varphi_3)$
and $M_2(\varphi_1,M_2(\varphi_2,\varphi_3))$. They are depicted on fig. \ref{fig:m2-m2}.
For a primary local operators we can write
\begin{equation}
\begin{split}
  f\circ M_2(M_2(\varphi_1,\varphi_2),\varphi_3)&= CF \cdot
  m_2^{\epsilon_2}(m_2^{\epsilon_1}(f\circ\varphi_1,f\circ\varphi_2),f\circ\varphi_3), \\
  f\circ M_2(\varphi_1,M_2(\varphi_2,\varphi_3))&= CF \cdot
  m_2^{\epsilon_2}(f\circ\varphi_1,m_2^{\epsilon_1}(f\circ\varphi_2,f\circ\varphi_3)),
\end{split}
\end{equation}
where we manifestly indicate the splittings $\epsilon_1$ and $\epsilon_2$ between the images of insertion points of the operators
that in the limit $\tau\to\infty$ behave as $\epsilon_1\sim e^{-2\tau+o(\tau)}$ and $\epsilon_2\sim e^{-\tau+o(\tau)}$.
Thus the maps $M_2$ being identical before the conformal mapping
seem to become different maps $m_2$ on the UHP gaining what we will call a hierarchy of the splittings:
the "inner" $m_2$ has $\epsilon_1$ as a splitting between their arguments, while the "outer" $m_2$ has splitting $\epsilon_2$ (see fig. \ref{fig:m2-m2-uhp}).
Of course, it doesn't mean that we have different maps on the UHP (they are the same, although are distorted by the conformal mapping),
but it implies that while doing calculation on the UHP we have to be careful when encounter a composition of the maps,
because composition results in the hierarchy on the~UHP.

\begin{figure}[ht!]
\begin{center}
  \begin{picture}(0,0)%
\includegraphics{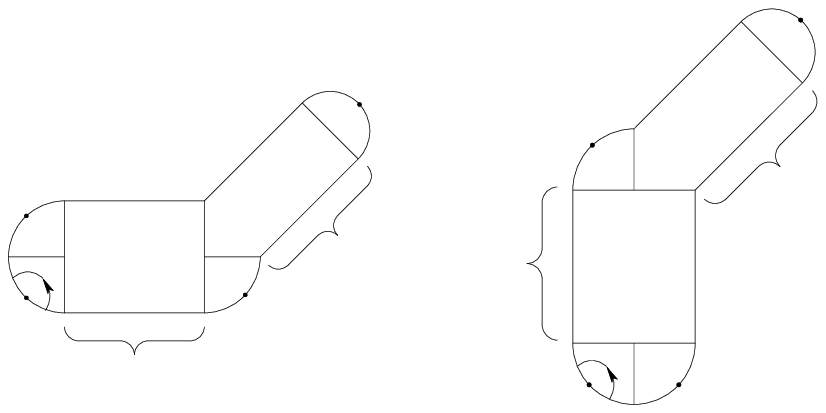}%
\end{picture}%
\setlength{\unitlength}{1243sp}%
\begingroup\makeatletter\ifx\SetFigFont\undefined%
\gdef\SetFigFont#1#2#3#4#5{%
  \reset@font\fontsize{#1}{#2pt}%
  \fontfamily{#3}\fontseries{#4}\fontshape{#5}%
  \selectfont}%
\fi\endgroup%
\begin{picture}(15154,6523)(-1454,-6515)
\put(13681,-4156){\makebox(0,0)[lb]{\smash{{\SetFigFont{17}{20.4}{\rmdefault}{\mddefault}{\updefault}{\color[rgb]{0,0,0}$\Bigr)$}%
}}}}
\put(7516,-4336){\makebox(0,0)[lb]{\smash{{\SetFigFont{17}{20.4}{\rmdefault}{\mddefault}{\updefault}{\color[rgb]{0,0,0}$-$}%
}}}}
\put(5086,-4966){\makebox(0,0)[lb]{\smash{{\SetFigFont{9}{10.8}{\rmdefault}{\mddefault}{\updefault}{\color[rgb]{0,0,0}$\varphi_3$}%
}}}}
\put(1981,-4426){\makebox(0,0)[lb]{\smash{{\SetFigFont{9}{10.8}{\rmdefault}{\mddefault}{\updefault}{\color[rgb]{0,0,0}$\int b$}%
}}}}
\put(9856,-6316){\makebox(0,0)[lb]{\smash{{\SetFigFont{9}{10.8}{\rmdefault}{\mddefault}{\updefault}{\color[rgb]{0,0,0}$\varphi_2$}%
}}}}
\put(8911,-4246){\makebox(0,0)[lb]{\smash{{\SetFigFont{9}{10.8}{\rmdefault}{\mddefault}{\updefault}{\color[rgb]{0,0,0}$s$}%
}}}}
\put(13546,-331){\makebox(0,0)[lb]{\smash{{\SetFigFont{9}{10.8}{\rmdefault}{\mddefault}{\updefault}{\color[rgb]{0,0,0}$\chi$}%
}}}}
\put(11566,-6361){\makebox(0,0)[lb]{\smash{{\SetFigFont{9}{10.8}{\rmdefault}{\mddefault}{\updefault}{\color[rgb]{0,0,0}$\varphi_3$}%
}}}}
\put(10666,-5731){\makebox(0,0)[lb]{\smash{{\SetFigFont{9}{10.8}{\rmdefault}{\mddefault}{\updefault}{\color[rgb]{0,0,0}$\int b$}%
}}}}
\put(6796,-1546){\makebox(0,0)[lb]{\smash{{\SetFigFont{9}{10.8}{\rmdefault}{\mddefault}{\updefault}{\color[rgb]{0,0,0}$\chi$}%
}}}}
\put(1216,-3166){\makebox(0,0)[lb]{\smash{{\SetFigFont{9}{10.8}{\rmdefault}{\mddefault}{\updefault}{\color[rgb]{0,0,0}$\varphi_1$}%
}}}}
\put(1306,-5101){\makebox(0,0)[lb]{\smash{{\SetFigFont{9}{10.8}{\rmdefault}{\mddefault}{\updefault}{\color[rgb]{0,0,0}$\varphi_2$}%
}}}}
\put(9631,-2131){\makebox(0,0)[lb]{\smash{{\SetFigFont{9}{10.8}{\rmdefault}{\mddefault}{\updefault}{\color[rgb]{0,0,0}$\varphi_1$}%
}}}}
\put(13096,-2941){\makebox(0,0)[lb]{\smash{{\SetFigFont{9}{10.8}{\rmdefault}{\mddefault}{\updefault}{\color[rgb]{0,0,0}$\tau$}%
}}}}
\put(3151,-5866){\makebox(0,0)[lb]{\smash{{\SetFigFont{9}{10.8}{\rmdefault}{\mddefault}{\updefault}{\color[rgb]{0,0,0}$s$}%
}}}}
\put(6391,-4021){\makebox(0,0)[lb]{\smash{{\SetFigFont{9}{10.8}{\rmdefault}{\mddefault}{\updefault}{\color[rgb]{0,0,0}$\tau$}%
}}}}
\put(-1439,-4291){\makebox(0,0)[lb]{\smash{{\SetFigFont{17}{20.4}{\rmdefault}{\mddefault}{\updefault}{\color[rgb]{0,0,0}$\int\limits_0^\tau ds\Bigl($}%
}}}}
\end{picture}%
  \caption{To the definition of $m_3$.}
  \label{fig:m3}
\end{center}
\end{figure}

Define $M_3$ this way
\footnote{From here on by $\{a,b\}$ we mean supercommutator.}
\begin{equation}\label{m3-mapping}
\begin{split}
    &M_3(\varphi_1(w_1),\varphi_2(w_2),\varphi_3(w_3))= \\
    &-t^{a_1}t^{a_2}t^{a_3} (-1)^{|\varphi_1|}\Bigl[
    \int_0^\tau ds\; e^{-\tau L_0} \varphi_1^{a_1}(w_1) e^{(s-\tau)L_0} \{b_{-1},\varphi_2^{a_2}(w_2)\}\varphi_3^{a_3}(w_3)\\
    &-\int_0^\tau ds\; e^{-\tau L_0} \varphi_3^{a_3}(w_3) e^{(s-\tau)L_0} \{b_{-1},\varphi_2^{a_2}(w_2)\}\varphi_1^{a_1}(w_1)
    \Bigr],
\end{split}
\end{equation}
where $|\varphi_1|$ is ghost number of $\varphi_1$
and $t^a$ is a generator of corresponding Lie algebra so
a local operator can be written as $\varphi=\varphi^at^a$ with $\varphi^a$ being a scalar.
Exponents act on all operators that stand to the right.
Fig. \ref{fig:m3} is a geometrical (visual) representation of the definition.
Thus we can see that $M_3$ is a homotopy between two ways of composing $M_2\circ M_2$.
This is a strict statement because, as we will see, $M_3$ satisfy the third relation \eqref{m2-m2} of $A_\infty$-algebra.

In general via mapping $M_3$ to the UHP we get something complicated because
doing conformal mapping we have to take into account transformation properties of the operators
we are acting on while the fields may have non-zero conformal dimension and even be non-primary.
But for a primary local operators of conformal dimension 0 we can write a simple expression
\begin{align}\label{m3-uhp-mapping}
  m_3^{\epsilon_1,\epsilon_2}(\varphi_1,\varphi_2,\varphi_3)(x)
  &= (-1)^{|\varphi_1|}\left[\varphi_1(x)\int_{x+\epsilon_1}^{x+\epsilon_2-\epsilon_1} dy\; \{b_{-1},\varphi_2(y)\}\;\varphi_3(x+\epsilon_2)\right]_x
\end{align}
with the splittings
\begin{equation}
  \epsilon_1\sim e^{-2\tau+o(\tau)},\quad \epsilon_2\sim e^{-\tau+o(\tau)}\quad \mbox{when } \tau\to\infty.
\end{equation}
As long as we don't have an explicit conformal mapping we don't know dependence of the splittings on $\tau$ exactly.
In actual calculations we need more precise relations between the splittings.
Thus we propose
\begin{equation}\label{hierarchy-m3-mapping}
  \epsilon_1/\epsilon_2\to0,\quad \epsilon_2^2/\epsilon_1\to0 \quad \mbox{when }  \tau\to\infty.
\end{equation}
So it is actually a prescription based on the mapping.

We see that in the case of primary local operators with conformal dimension 0
the mapping to the UHP results only in the hierarchy of the splittings and we don't need to know the mapping exactly.

The first three maps $M_k$ satisfy first three quadratic relations of $A_\infty$-algebra:
\begin{align}
  &M_1^2 = 0, \\
  &M_1 M_2(A,B) = M_2(M_1(A),B)+(-1)^{|A|}M_2(A,M_1(B)), \\ \nonumber
  &M_2(A,M_2(B,C))-M_2(M_2(A,B),C) = \label{m2-m2} \\ \nonumber
    &=M_1 M_3(A,B,C) \\
    &\;\;\;+M_3(M_1(A),B,C)+(-1)^{|A|}M_3(A,M_1(B),C)+(-1)^{|A|+|B|}M_3(A,B,M_1(C)).
\end{align}
Indeed, this relations may be trivially checked using definitions \eqref{m2-mapping} and \eqref{m3-mapping}.

Conformal mapping to the UHP does not spoil this property,
but again one should be careful when doing compositions.
Thus writing similar conditions on the UHP
we manifestly indicate the splittings:
\begin{align}
  &m_1^2 = 0, \\
  &m_1 m_2^\epsilon(A,B) = m_2^\epsilon(m_1(A),B)+(-1)^{|A|}m_2^\epsilon(A,m_1(B)), \\ \nonumber
  &m_2^{\epsilon_2}(A,m_2^{\epsilon_1}(B,C))-m_2^{\epsilon_2}(m_2^{\epsilon_1}(A,B),C) = \label{m2-m2-UHP} \\ \nonumber
    &=m_1 m_3^{\epsilon_1,\epsilon_2}(A,B,C) \\
    &\;\;\;+m_3^{\epsilon_1,\epsilon_2}(m_1(A),B,C)+(-1)^{|A|}m_3^{\epsilon_1,\epsilon_2}(A,m_1(B),C)+(-1)^{|A|+|B|}m_3^{\epsilon_1,\epsilon_2}(A,B,m_1(C)).
\end{align}
See Appendix \ref{section:quad-rel} for manifest verification of this relations.

We know that the map $M_3$ is a homotopy between two possible compositions of $M_2\circ M_2$.
The map $M_3$ is an integral over an edge of polytope called associahedra \cite{Stasheff-1, A-inf}.
Following this route we can define each higher $M_k$
as an integral over corresponding $(k-2)$-dimensional face of the associahedra.
Having in mind this homotopical definition we propose
that the rest of conditions of $A_\infty$-algebra are satisfied.

We still have $m_k=f\circ M_k$, but let us work out this definition on the UHP.
In order to define $m_k$ we take all ways of composing $k-1$ instances of $m_2$ into a map $\mathcal H^{\otimes k}\to\mathcal H$.
Each composition is an OPE of $k$ operators inserted at the real line.
Insertion points of the operators are governed by hierarchy of splittings $\epsilon_1,...,\epsilon_{k-1}$ and structure of concrete composition.
The coordinates of insertion points define a point in $\rr^k$.
So for each way of composing $k-1$ instances of $m_2$ we have a point in $\rr^k$.
There is $(k-2)$-dimensional face $D_k$ of associahedra with these points being vertices.
Define map $m_k$
on a primary local operators of conformal dimension 0
as the following integral
\begin{equation}\label{mk-uhp-mapping}
\begin{split}
  &m_k^{\epsilon_1,...,\epsilon_{k-1}}(\varphi_1,...,\varphi_k)(x_1)=\\
  &=(-1)^{(k-2)|\varphi_1|}\ldots(-1)^{|\varphi_{k-2}|}
  \Bigl[\varphi_1(x_1)\int_{D_k}dx_2 ... dx_{k-1}\; \{b_{-1},\varphi_2(x_2)\} ...
  \{b_{-1},\varphi_{k-1}(x_{k-1})\} \; \varphi_k(x_k)\Bigr]_{x_1}
\end{split}
\end{equation}
From the definition it follows that the boundary of $D_k$ is union of $(k-3)$-dimensional faces,
each of them being integration domain for an instance of $m_{k-1}$.
Using decent relations we can obtain quadratic relations of $A_\infty$-structure from the structure of associahedra.
Schematic derivation is presented below (we suppress all signs and splittings for readability):
\begin{equation}
\begin{split}
  &Q m_k(\varphi_1,...,\varphi_k)-m_k(Q\varphi_1,...,\varphi_k)-...-m_k(\varphi_1,...,Q\varphi_k)\\
  &=\int_{D_k} \varphi_1(x_1)[L_0,\varphi_2(x_2)] ...\{b_{-1},\varphi_{k-1}(x_{k-1})\} \; \varphi_k(x_k) +...+\\
  &\quad \int_{D_k} \varphi_1(x_1)\{b_{-1},\varphi_2(x_2)\} ...[L_0,\varphi_{k-1}(x_{k-1})] \; \varphi_k(x_k)\\
  &=\int_{D_k} \varphi_1(x_1)\partial \varphi_2(x_2) ...\{b_{-1},\varphi_{k-1}(x_{k-1})\} \; \varphi_k(x_k)+...+\\
  &\quad\int_{D_k} \varphi_1(x_1)\{b_{-1},\varphi_2(x_2)\} ...\partial \varphi_{k-1}(x_{k-1}) \; \varphi_k(x_k)\\
  &=\sum_{n=1}^{k-2}\bigl[\pm m_{n+1}(\varphi_1,...,\varphi_n,m_{k-n}(\varphi_{n+1},...,\varphi_k))\pm...\pm m_{n+1}(m_{k-n}(\varphi_1,...,\varphi_{k-n}),\varphi_{k-n+1},...,\varphi_k))\bigr]
\end{split}
\end{equation}
Example of $m_k$ for $k=3$ was given above. 
Later we will present example for $k=4$ where $D_4$ is a region defined by relations \eqref{D4}
with the hierarchy of splittings \eqref{hierarchy-m4}.

In conclusion of the section let us recall that we have introduced the maps $M_k$ that depend
on a parameter $\tau$. These maps form $A_\infty$-algebra thus depending on the parameter too.
Later in order to reproduce the Yang-Mills equation we will take a limit $\tau\to\infty$.
Under conformal mapping we have obtained the maps $m_k$ that depends on a set of a parameters (which depend on $\tau$) -- point splittings.
In the limit $\tau\to\infty$ the splittings go to zero with their own pace
that is governed by what we called the hierarchy of the splittings.
The hierarchy appears via the conformal mapping of the maps $M_k$ to the UHP.
We repeat our note that in actual calculations we use proposed hierarchy of splittings based on
actual hierarchy, but it is not proved that these hierarchies are equal.

\subsection{The equation}
Having defined maps let us consider the homotopical Maurer-Cartan equation.
By the hMC equation we mean a generalization of common Maurer-Cartan equation to $A_\infty$-algebras.
In this section we study properties of the hMC equation written in terms of $M_k$,
but all this properties are also true for the equation written in terms of $m_k$,
because they are related by the conformal mapping.
The equation associated with the maps $M_k$ is
\begin{equation}\label{hMC}
M_1(V)+M_2(V,V)-M_3(V,V,V)+...=0
\end{equation}
with $V$ being a ghost number 1 local operator.

Map $M_k$ has ghost number $2-k$, so by applying it to $V$ we get
a local operator $M_k(V,...,V)$ with ghost number $2$.

As long as the maps $M_k$ form $A_\infty$-algebra the symmetry of the equation is
\begin{equation}
\begin{split}\label{delta-hMC}
  \delta V&=M_1(H)+M_2(V,H)-M_2(H,V)\\
  &-M_3(H,V,V)+M_3(V,H,V)-M_3(V,V,H)+...,
\end{split}
\end{equation}
where parameter of the symmetry $H$ is ghost number 0 local operator and
ellipsis stands for the subsequent maps where we have substituted one $H$ instead of $V$ in a different places in $M_k$ keeping track of the sign.

Thus we consider a local operator from $\Omega^1$ as a variable.
Later in order to obtain the Yang-Mills equation we specialize to the local operators defined by means of
a field $A_\mu$ which we will call a parametrization.
The equation on this operator is in $\Omega^2$.
The parameter of symmetry of the equation is in $\Omega^0$.

In attempt to solve the hMC equation with a local operator $V$ in a particular parametrization~$A_\mu$
one may encounter an obstruction.
Obstruction is a cohomology of the space $\Omega^2$ of ghost number 2 local operators.
Thus for the local operator to be a solution of the hMC equation
the parametrization must satisfy an equation -- an obstruction vanishing condition.
This equation has a symmetry that is inherited from the symmetry $\delta V$ of the hMC equation.

In order to obtain the Yang-Mills equation from the hMC equation we have to reproduce structure of the Yang-Mills equation
from the obstruction vanishing condition and the symmetry $\delta\mathcal A_\mu$ of the Yang-Mills equation from  the symmetry $\delta V$ of the hMC equation.
As a result, we obtain (possibly) nonlinear dependence $\mathcal A_\mu=\mathcal A_\mu(A)$
of a gauge potential $\mathcal A_\mu$ on the parametrization $A_\mu$ that may lead to identification of $\ap$-correction to the Yang-Mills equation.

In general all the maps $M_k$ depend on the parameter $\tau$.
We match the hMC and the Yang-Mills equations in a limit $\tau\to\infty$.
Thus we are only left with expressions that don't depend on the parameter and with singular ones.
But we don't expect appearance of the parameter in the obstruction vanishing condition.
So we expect singular in the parameter terms to be exact.

As long as we will be doing all calculation on the UHP
it is sensible to reformulate this procedure in terms of the splittings.
Namely, we match the hMC and the Yang-Mills equations in a limit when all the splittings go to zero,
although each with their own pace governed by a hierarchy of splittings.
Conditions \eqref{hierarchy-m3-mapping} provides us with the first example of relations in the hierarchy.
Thus we are only left with expressions that don't depend on the parameters and with singular ones
with latter expected to be exact.

To realize this approach let us start with examination of a structure of the Yang-Mills equation.
The equation can be rewritten as follows
\begin{equation}\label{alg-YM}
  \mathrm{Max}(\mathcal A)+\mathrm{YM}^2(\mathcal A)+\mathrm{YM}^3(\mathcal A)=0
\end{equation}
with $\mathrm{Max},\mathrm{YM}^2$ and $\mathrm{YM}^3$ being linear, quadratic and cubic in gauge potential $\mathcal A_\mu$
parts of the Yang-Mills equation correspondingly
\begin{align}
  \mathrm{Max}(\mathcal A_\mu)&=
    \Box\mathcal A_\mu-\partial_\mu\partial_\nu\mathcal A_\nu, \label{Max-part} \\
  \mathrm{YM}^2(\mathcal A_\mu)&=
    \partial_\nu[\mathcal A_\nu,\mathcal A_\mu]+
    [\mathcal A_\nu,\partial_\nu\mathcal A_\mu-\partial_\mu\mathcal A_\nu], \label{YM2-part} \\
  \mathrm{YM}^3(\mathcal A_\mu)&=[\mathcal A_\nu,[\mathcal A_\nu,\mathcal A_\mu]]. \label{YM3-part}
\end{align}
The symmetry of \eqref{alg-YM} is
\begin{equation}
  \delta\mathcal  A_\mu=\partial_\mu\varepsilon+[\mathcal A_\mu,\varepsilon]
\end{equation}
with $\varepsilon$ being an infinitesimal parameter of the symmetry.
Seeking solution in a perturbation series in a coupling $t$
\begin{align}
  \mathcal A_\mu&=t\mathcal A_\mu^{(1)}+t^2\mathcal A_\mu^{(2)}+..., \\
  \varepsilon&=t\varepsilon^{(1)}+t^2\varepsilon^{(2)}+...
\end{align}
we get the following equations
\begin{align}
  &t^1:\; \mathrm{Max}(\mathcal A_\mu^{(1)})=0, \label{YM1} \\
  &t^2:\; \mathrm{Max}(\mathcal A_\mu^{(2)})+\mathrm{YM}^2(\mathcal A_\mu^{(1)})=0, \label{YM2} \\ \nonumber
  &t^3:\; \mathrm{Max}(\mathcal A_\mu^{(3)})+\mathrm{YM}^3(\mathcal A_\mu^{(1)}) \\ \nonumber
  &\quad\quad +2[\mathcal A_\nu^{(1)},\partial_\nu\mathcal A_\mu^{(2)}]
	      +2[\mathcal A_\nu^{(2)},\partial_\nu\mathcal A_\mu^{(1)}]
	      -[\mathcal A_\nu^{(1)},\partial_\mu\mathcal A_\nu^{(2)}]
	      -[\mathcal A_\nu^{(2)},\partial_\mu\mathcal A_\nu^{(1)}]\\
  &\quad\quad -[\mathcal A_\mu^{(1)},\partial_\nu\mathcal A_\nu^{(2)}]
	      -[\mathcal A_\mu^{(2)},\partial_\nu\mathcal A_\nu^{(1)}]=0. \label{YM3}
\end{align}
and so on. Their symmetries are
\begin{align}
  \delta\mathcal  A^{(1)}_\mu &= \partial_\mu\varepsilon^{(1)}, \\
  \delta\mathcal  A^{(2)}_\mu &= \partial_\mu\varepsilon^{(2)}+[\mathcal A_\mu^{(1)},\varepsilon^{(1)}], \\
  \delta\mathcal  A^{(3)}_\mu &= \partial_\mu\varepsilon^{(3)}+[\mathcal A_\mu^{(1)},\varepsilon^{(2)}]+[\mathcal A_\mu^{(2)},\varepsilon^{(1)}] \label{YM-symm-3}
\end{align}
and so on.

Next let us find a similar structures in the hMC equation and its symmetry.
To do this we again take an attempt to find a solution to the hMC equation \eqref{hMC} as a perturbation series in the coupling $t$
\begin{equation}
  V=tV_1+t^2V_2+...,
\end{equation}
where each $V_k$ is also a series in $\ap$. This way we obtain the following equations
\begin{align}
  t^1: \;\;& M_1(V_1)=0, \label{hMC1} \\
  t^2: \;\;& M_1(V_2)+M_2(V_1,V_1)=0, \label{hMC2} \\
  t^3: \;\;& M_1(V_3)+M_2(V_1,V_2)+M_2(V_2,V_1)-M_3(V_1,V_1,V_1)=0 \label{hMC3}
\end{align}
and so on.
Corresponding equations on the UHP are
\begin{align}
  t^1: \;\;& m_1(V_1)=0, \label{hMC1-uhp} \\
  t^2: \;\;& m_1(V_2)+m_2^\epsilon(V_1,V_1)=0, \label{hMC2-uhp} \\
  t^3: \;\;& m_1(V_3)+m_2^{\epsilon_2}(V_1,V_2)+m_2^{\epsilon_2}(V_2,V_1)-m_3^{\epsilon_1,\epsilon_2}(V_1,V_1,V_1)=0 \label{hMC3-uhp}
\end{align}
and so on.
We will be solving these particular equations in section \ref{section:YM}.
As long as the local operator becomes a series so do the parametrization $A_\mu=\sum_{k\ge1}t^kA_\mu^{(k)}$
and the symmetry parameter $H=\sum_{k\ge1}t^kH^{(k)}$.

Let us examine some properties that do not depend on a particular choice of parametrization.
The first equation is
\begin{equation}
  M_1(V_1)=0.
\end{equation}
It has a symmetry that is simply shift by an exact element
\begin{equation}
  \delta V_1=M_1(H^{(1)}).
  \label{dV1}
\end{equation}

The second equation is
\begin{equation}
  M_1(V_2)+M_2(V_1,V_1)=0.
\end{equation}
It has more complicated symmetry containing apart from an exact element $m_2$-commutator
\begin{align}
  \delta V_1 &= M_1(H^{(1)}), \\
  \delta V_2 &= M_1(H^{(2)}) +M_2(V_1,H^{(1)})-M_2(H^{(1)},V_1).
\end{align}
Similar reasoning could be done for a subsequent equations, but we better switch to a concrete example.

\section{Emergence of the Yang-Mills equation}\label{section:YM}


In this section we present derivation of the Yang-Mills equation from the hMC equation \eqref{hMC1-uhp}-\eqref{hMC3-uhp}.
To do this we start with the following parametrization of the local operator in the first order in the coupling~$t$
\begin{equation}\label{v1}
  V_1=cA_\mu^{(1)}\partial X^\mu-\frac\ap4\partial c\partial_\mu A_\mu^{(1)},
\end{equation}
where $A_\mu^{(1)}=A_\mu^{(1)}(X)$ is a functions on the target space.
For $V_1$ we have
\begin{align}
  Q(V_1)&=\frac\ap4c\partial c\Bigl[ \Box A_\mu^{(1)} -\partial_\mu\partial_\nu A_\nu^{(1)}\Bigr]\partial X^\mu,
\end{align}
so in this parametrization equation \eqref{hMC1-uhp} takes form
\begin{equation}
  \frac\ap4c\partial c\Bigl[ \Box A_\mu^{(1)} -\partial_\mu\partial_\nu A_\nu^{(1)}\Bigr]\partial X^\mu=0.
\end{equation}
We see that $V_1$ solves \eqref{hMC1-uhp} when parametrization $A_\mu^{(1)}$ satisfy Maxwell equation.
As we discussed earlier \eqref{dV1} this equation has a symmetry which is a shift by an exact element $V_1\to V_1+Q(H^{(1)})$.
The form of the exact element is
\begin{equation}
  Q(H^{(1)})=c\partial_\mu H^{(1)}\partial X^\mu-\frac\ap4\partial c\Box H^{(1)}.
\end{equation}
This shift naturally generates Maxwell gauge transformation for \eqref{v1}
\begin{equation}
  \delta A_\mu^{(1)}=\partial_\mu H^{(1)}.
\end{equation}
Thus the symmetry of the hMC equation at the first order in coupling $t$ implies the symmetry of the Maxwell (or linear part of the Yang-Mills) equation.

So at the first order in $t$ we have the following identifications of
gauge potential $\mathcal A$ and parameter of gauge transformation $\varepsilon$ with
parametrizations of local operators:
\begin{align}
  \mathcal A_\mu^{(1)}&=A_\mu^{(1)}, \\
  \varepsilon^{(1)}&=H^{(1)}
\end{align}
together with the Maxwell equation
\begin{equation}
  \mathrm{Max}(\mathcal A_\mu^{(1)})=0
\end{equation}
and its gauge symmetry
\begin{equation}
  \delta\mathcal A_\mu^{(1)}=\partial_\mu \varepsilon^{(1)}.
\end{equation}
Due to a form of the parametrization, structure of the equation and the symmetry
there is no way for $\ap$-correction to appear at this stage.


Situation changes when we go to the second order in the coupling $t$.
In order to find an obstruction we calculate $m_2^\epsilon(V_1,V_1)$ in \eqref{hMC2-uhp} up to terms $O(\ap^3)$.
The obstruction may be written in the form
\begin{equation}\label{YM-corr}
\begin{split}
  \frac\ap4\Bigl[\mathrm{Max}\left(A_\mu^{(2)}-\frac\ap2A_\alpha^{(1)}\partial_\mu A_\alpha^{(1)}\right)+\mathrm{YM}^2(A_\mu^{(1)}) \\
  +\ap\left(
    [\partial_\alpha A_\beta^{(1)},\partial_\mu F_{\alpha\beta}^{(1)}]
    +[\mathrm{Max}(A_\alpha^{(1)}),\partial_\mu A_\alpha^{(1)}]
  \right)\\
  -\frac\ap2\Bigl[
    2[\partial_\alpha A_\beta^{(1)},\partial_\alpha\partial_\beta A_\mu^{(1)}]
    -[\partial_\alpha A_\beta^{(1)},\partial_\alpha\partial_\mu A_\beta^{(1)}]
    -[\partial_\alpha A_\mu^{(1)},\partial_\alpha\partial_\beta A_\beta^{(1)}]\Bigr]\log\epsilon^2 +O(\ap^2)\Bigr]\partial X^\mu=0.
\end{split}
\end{equation}
with $F_{\alpha\beta}^{(1)}=\partial_\alpha A_\beta^{(1)}-\partial_\beta A_\alpha^{(1)}$ being a prototype of a linear part of field strength.

The second line of \eqref{YM-corr} coincides with $\ap \partial_\alpha[ F_{\mu\beta}^{(1)},F_{\beta\alpha}^{(1)}]$
that is a $\partial^3A^2$-part of a variation of $\mathrm{Tr}(F^3)$ from the non-abelian generalization of Born-Infeld action
\cite{Tseytlin-BI,YM-from-OSFT}
modulo Maxwell equation on $A_\mu^{(1)}$.
Namely our result differs from the variation of $\mathrm{Tr}(F^3)$ by $[\partial_\beta A_\mu^{(1)},\mathrm{Max}(A_\beta^{(1)})]$.
Thus instead of the second line in \eqref{YM-corr} we will write $\ap \partial_\alpha[F_{\mu\beta}^{(1)},F_{\beta\alpha}^{(1)}]$.

As we can see this obstruction does depend on the splitting via $\log\epsilon$.
But structure with $\log\epsilon$ is similar to the $\mathrm{YM}^2$-part \eqref{YM2-part}.
This fact allow us to do a redefinition below that will help us to deal with $\log\epsilon$-terms.

For $V_2$ we have the following expression
\begin{align}\label{V2}
  V_2&=
    cA_\mu^{(2)}\partial X^\mu-\frac\ap4\partial c\partial_\mu A_\mu^{(2)}
    +\frac\ap2 \left(\frac{c}{\epsilon}+\frac{\partial c}{2}\right)A_\mu^{(1)}A_\mu^{(1)}
    +O(\ap^2)
\end{align}
Its symmetry can be found this way. A variation of \eqref{hMC2-uhp} is
\begin{equation}
  Q(\delta V_2)+m_2^\epsilon(\delta V_1,V_1)+m_2^\epsilon(V_1,\delta V_1)=0.
\end{equation}
As long as $\delta V_1=QH^{(1)}$ and due to \eqref{hMC1-uhp} we can write
\begin{equation}
  Q\Bigl(\delta V_2-m_2^\epsilon(V_1,H^{(1)})+m_2^\epsilon(H^{(1)},V_1)\Bigr)=0
\end{equation}
so
\begin{equation}
  \delta V_2=QH^{(2)}+m_2^\epsilon(V_1,H^{(1)})-m_2^\epsilon(H^{(1)},V_1).
\end{equation}
Doing some math we arrive to
\begin{equation}
\begin{split}
  \delta V_2
  &= c\Bigl[\partial_\mu H^{(2)}+[A_\mu^{(1)},H^{(1)}]\Bigr]\partial X^\mu
    -\frac\ap4 \partial c\partial_\mu\Bigl[\partial_\mu H^{(2)} +[A_\mu^{(1)},H^{(1)}]\Bigr]\\
  &+\frac\ap2\left(\frac c\epsilon+\frac{\partial c}2\right)\{A_\mu^{(1)},\partial_\mu H^{(1)}\}
  +\delta_\mathrm{Max}\left( \frac\ap2 cA_\alpha^{(1)}\partial_\mu A_\alpha^{(1)} \right)\partial X^\mu \\
  &-\frac\ap2 c[\partial_\alpha A_\mu^{(1)},\partial_\alpha H^{(1)}]\partial X^\mu \log\epsilon^2
    +O(\ap^2),
\end{split}
\end{equation}
which besides a variation of \eqref{V2} contains some other terms and $\delta_\mathrm{Max}A_\mu^{(1)}=\partial_\mu H^{(1)}$.

The following redefinition
\begin{equation}
  \tilde A_\mu^{(1)}= \left(1-\frac\ap4\log\epsilon^2\Box\right)A_\mu^{(1)},
\end{equation}
where $\Box=\eta^{\mu\nu}\partial_\mu\partial_\nu$ is d'Alambertian on the target space
helps us to deal with $\log\epsilon$.
The equation becomes
\begin{equation}\label{eq-after-redef}
\begin{split}
  \left[1+\frac\ap4\log\epsilon^2 \Box\right]\left[
  \mathrm{Max}\left(\tilde A_\mu^{(2)}
  -\frac\ap2\tilde A_\alpha^{(1)}\partial_\mu \tilde A_\alpha^{(1)}\right)+
  \mathrm{YM}^2(\tilde A_\mu^{(1)})+\ap \partial_\alpha[\tilde F_{\mu\beta}^{(1)},\tilde F_{\beta\alpha}^{(1)}]
  \right] 
  +O(\ap^2)=0.
\end{split}
\end{equation}
As long as $1+\frac\ap4\log\epsilon^2\Box$ is invertible (at least perturbatively) we may omit it.

Doing more contractions in $m_2^\epsilon(V_1,V_1)$ will only have the effect of the redefinition (see App. \ref{section:redef}).
The full redefinition is 
\begin{equation}\label{A-redef-1}
  \tilde A_\mu^{(1)}= \exp\left(-\frac\ap4\log\epsilon^2\Box\right)A_\mu^{(1)},
\end{equation}
which is similar to redefinition made in \cite{YM-from-OSFT} while deriving the Yang-Mills action from the bosonic open string field theory action.
Also, the Yang-Mills equation was obtained form the open superstring field theory in \cite{YM-from-Super}.

After this redefinition the symmetry becomes
\begin{equation}\label{symm-after-redef}
\begin{split}
  \delta V_2
  &= c\Bigl[\partial_\mu H^{(2)}+[\tilde A_\mu^{(1)},H^{(1)}]\Bigr]\partial X^\mu
    -\frac\ap4 \partial c\partial_\mu\Bigl[\partial_\mu H^{(2)} +[\tilde A_\mu^{(1)},H^{(1)}]\Bigr]\\
  &+\frac\ap2\left(\frac c\epsilon+\frac{\partial c}2\right)\{\tilde A_\mu^{(1)},\partial_\mu H^{(1)}\}
  +\delta_\mathrm{Max}\left( \frac\ap2 c\tilde A_\alpha^{(1)}\partial_\mu \tilde A_\alpha^{(1)} \right)\partial X^\mu
    +O(\ap^2).
\end{split}
\end{equation}

The form of the equation \eqref{eq-after-redef} and its symmetry \eqref{symm-after-redef} forces us to do
the following identifications at the second order in the coupling $t$
\begin{align}\label{A-redef-2}
  \mathcal A_\mu^{(2)} &= \tilde A_\mu^{(2)}-\frac\ap2\tilde A_\alpha^{(1)}\partial_\mu \tilde A_\alpha^{(1)}, \\
  \varepsilon^{(2)}&=H^{(2)}.
\end{align}
In this terms quadratic part of the Yang-Mills equation is
\begin{equation}\label{alpha-corr}
  \mathrm{Max}(\mathcal A_\mu^{(2)})+\mathrm{YM}^2(\mathcal A_\mu^{(1)})
  +\ap \partial_\alpha[\mathcal F_{\mu\beta}^{(1)},\mathcal F_{\beta\alpha}^{(1)}]+O(\ap^2)=0,
\end{equation}
where
\begin{equation}
  \mathcal F_{\alpha\beta}^{(1)}=
    \partial_\alpha\mathcal A_\beta^{(1)}-\partial_\beta\mathcal A_\alpha^{(1)}
\end{equation}
is a linear part of field strength
and the gauge symmetry is
\begin{equation}
\begin{split}
  \delta\mathcal A_\mu^{(1)}&=\partial_\mu \varepsilon^{(1)},\\
  \delta\mathcal A_\mu^{(2)}&=\partial_\mu \varepsilon^{(2)}+[\mathcal A_\mu^{(1)},\varepsilon^{(1)}].
\end{split}
\end{equation}
So we see that using approach based on solving the hMC equation we have reproduced quadratic part of the Yang-Mills equation together
with $\ap$-correction that is linear in $\ap$ which arises from the $\mathrm{Tr}(\mathcal F^3)$ term
in the non-abelian generalization of Born-Infeld action.

Thus we are able to propose a general method for calculating $\ap$-corrections to the Yang-Mills equation from the hMC equation.
One should calculate a result of each map up to desired order in $\ap$ or in other words do a proper number of contractions.
In general one should keep regular terms in the splittings which may cancel out singular terms,
however in our calculations they have never contributed to the equation.
Then exact terms may be absorbed by $V_k$ and non-exact terms form an obstruction that
after proper redefinition of the parametrization provide us with the equation and desired $\ap$-corrections to it.


Let us get a cubic term of the Yang-Mills equation form the hMC equation.
In order to do this we have to find an obstruction to the equation
\begin{equation}\label{KS-3-order}
  Q(V_3)+m_2^{\epsilon_2}(V_1,V_2)+m_2^{\epsilon_2}(V_2,V_1)-m_3^{\epsilon_1,\epsilon_2}(V_1,V_1,V_1)=0
\end{equation}
with $m_3$ given by \eqref{m3-uhp-mapping}.
In order to calculate $m_2^{\epsilon_2}(V_2,V_1)$ we note that $V_2$ consists of two parts:
the first has zero conformal dimension thus doesn't transform,
the second part $cA^2/\epsilon$ with $\epsilon\sim e^{-\tau+o(\tau)}$ has nonzero dimension and under the mapping
becomes $cA^2/\epsilon_1$ with $\epsilon_1\sim e^{-2\tau+o(\tau)}$.
Result of each map
\begin{align} \nonumber
  &m_3^{\epsilon_1,\epsilon_2}(V_1,V_1,V_1) = \\
    &\quad=\frac\ap2c\partial c\left[ A_\nu^{(1)} A_\mu^{(1)} A_\nu^{(1)} + (A_\nu^{(1)}A_\nu^{(1)}A_\mu^{(1)} +A_\mu^{(1)} A_\nu^{(1)}A_\nu^{(1)})
    \left(\frac{\epsilon_2}{\epsilon_1}-1\right) \right]\partial X^\mu \\ \nonumber
    &\quad\quad+O(\mbox{splittings})+O(\ap^2), \nonumber
\end{align}
\begin{align}\nonumber
  &m_2^{\epsilon_2}(V_2,V_1) = +\frac\ap2c\partial c\left[ \frac{\epsilon_2}{\epsilon_1}-\frac12\right]A_\nu^{(1)}A_\nu^{(1)}A_\mu^{(1)}\partial X^\mu\\ \nonumber
    &\quad+\frac\ap4\left[ 2A_\nu^{(2)}\partial_\nu A_\mu^{(1)} -2\partial_\nu A_\mu^{(2)}A_\nu^{(1)}
    +\partial_\mu A_\nu^{(2)} A_\nu^{(1)} -A_\nu^{(2)}\partial_\mu A_\nu^{(1)}
    +\partial_\nu A_\nu^{(2)} A_\mu^{(1)} -A_\mu^{(2)}\partial_\nu A_\nu^{(1)}\right] \\
    &\quad+Q\left[-\frac\ap2\left(\frac{c}{\epsilon_2}+\frac{\partial c}2\right)A_\mu^{(2)}A_\mu^{(1)}\right]+O(\mbox{splittings})+O(\ap^2),
\end{align}
\begin{align}\nonumber
  &m_2^{\epsilon_2}(V_1,V_2) = +\frac\ap2c\partial c\left[ \frac{\epsilon_2}{\epsilon_1}-\frac12\right]A_\mu^{(1)} A_\nu^{(1)}A_\nu^{(1)}\partial X^\mu\\ \nonumber
    &\quad+\frac\ap4\left[ 2A_\nu^{(1)}\partial_\nu A_\mu^{(2)} -2\partial_\nu A_\mu^{(1)}A_\nu^{(2)}
    +\partial_\mu A_\nu^{(1)} A_\nu^{(2)} -A_\nu^{(1)}\partial_\mu A_\nu^{(2)}
    +\partial_\nu A_\nu^{(1)} A_\mu^{(2)} -A_\mu^{(1)}\partial_\nu A_\nu^{(2)}\right]\\
    &\quad+Q\left[-\frac\ap2\left(\frac{c}{\epsilon_2}+\frac{\partial c}2\right)A_\mu^{(1)}A_\mu^{(2)}\right]+O(\mbox{splittings})+O(\ap^2),
\end{align}
where $O(\mathrm{splittings})$ is a term which vanishes as splittings go to zero according to \eqref{hierarchy-m3-mapping}.
Though each term diverges, their sum is finite
\begin{equation}\label{3-pt-hMC}
\begin{split}
  & m_2^{\epsilon_2}(V_1,V_2)+m_2^{\epsilon_2}(V_2,V_1)-m_3^{\epsilon_1,\epsilon_2}(V_1,V_1,V_1) \\
  & =\frac\ap4c\partial c\Bigl[ \mathrm{Max}(A^{(3)})+ \mathrm{YM}^2(A^{(2)},A^{(1)})+\mathrm{YM}^2(A^{(1)},A^{(2)})
  +\mathrm{YM}^3(A^{(1)}) \Bigr]\partial X^\mu \\
  &+Q\left[-\frac\ap2\left(\frac{c}{\epsilon_2}+\frac{\partial c}2\right)\{A_\mu^{(1)},A_\mu^{(2)}\}\right]+O(\ap^2).
\end{split}
\end{equation}
Thus
\begin{align}
  V_3 =cA_\mu^{(3)}\partial X^\mu-\frac\ap4\partial c\partial_\mu A_\mu^{(3)}+
    \frac{\ap}2\left(\frac{c}{\epsilon_2}+\frac{\partial c}2\right)\{A_\mu^{(1)},A_\mu^{(2)}\}+O(\ap^2).
\end{align}

The symmetry $\delta V_3$ contains besides expected terms that provide us with \eqref{YM-symm-3},
a few terms that go to a redefinitions in a similar to \eqref{A-redef-1} and \eqref{A-redef-2} way,
but in order to obtain cubic term of the Yang-Mills equation the 0-th order in $\ap$ is enough.
Thus at the third order in coupling $t$ we have the following identifications
\begin{align}
  \mathcal A_\mu^{(3)}&=A_\mu^{(3)}+O(\ap), \\
  \varepsilon^{(3)}&=H^{(3)}
\end{align}
together with a cubic part \eqref{YM3} of the Yang-Mills equation
\begin{equation}
\begin{split}
  &\mathrm{Max}(\mathcal A_\mu^{(3)})+\mathrm{YM}^3(\mathcal A_\mu^{(1)}) \\
  &\quad +2[\mathcal A_\nu^{(1)},\partial_\nu\mathcal A_\mu^{(2)}]
	      +2[\mathcal A_\nu^{(2)},\partial_\nu\mathcal A_\mu^{(1)}]
	      -[\mathcal A_\nu^{(1)},\partial_\mu\mathcal A_\nu^{(2)}]
	      -[\mathcal A_\nu^{(2)},\partial_\mu\mathcal A_\nu^{(1)}]\\
  &\quad -[\mathcal A_\mu^{(1)},\partial_\nu\mathcal A_\nu^{(2)}]
	      -[\mathcal A_\mu^{(2)},\partial_\nu\mathcal A_\nu^{(1)}]+O(\ap)=0
\end{split}
\end{equation}
and its gauge symmetry
\begin{equation}
\begin{split}
  \delta\mathcal A_\mu^{(1)}&=\partial_\mu \varepsilon^{(1)},\\
  \delta\mathcal A_\mu^{(2)}&=\partial_\mu \varepsilon^{(2)}+[\mathcal A_\mu^{(1)},\varepsilon^{(1)}],\\
  \delta\mathcal A_\mu^{(3)}&=\partial_\mu \varepsilon^{(3)}
    +[\mathcal A_\mu^{(2)},\varepsilon^{(1)}]+[\mathcal A_\mu^{(1)},\varepsilon^{(2)}].
\end{split}
\end{equation}
It is not hard to show that the hMC equation will reproduce the Yang-Mills equation order by order in the coupling $t$.
Moreover, in each order in $t$ only a few terms in the hMC equation will contribute.
This happens because a contribution of the maps $m_k$ with $k\ge 4$ are of $\ap^2$ order.
For example let us calculate the following map
\begin{align}
  & m_4^{\tilde\epsilon_1,\tilde\epsilon_2,\tilde\epsilon_3}(V_1,V_1,V_1,V_1)= \nonumber \\
  & -\left(cA_\mu\partial X^\mu-\frac\ap4\partial c\partial_\mu A_\mu\right)(x)
    \int_{D_4} dydz \; A_\nu\partial X^\nu(y)
    A_\alpha\partial X^\alpha(z)
    \left(cA_\beta\partial X^\beta-\frac\ap4\partial c\partial_\beta A_\beta\right)(w).
\end{align}
Here $D_4$ is a two-dimensional domain defined by the following constraints,
\begin{equation}\label{D4} 
\begin{split}
  &y-x>\tilde\epsilon_1,\quad
  z-y>\tilde\epsilon_1,\quad
  w-z>\tilde\epsilon_1,\\
  &z-x>\tilde\epsilon_2,\quad
  w-y>\tilde\epsilon_2
\end{split}
\end{equation}
where $x$, $w$ are fixed with $w-x=\tilde\epsilon_3$ and $y$, $z$ are coordinates describing $D_4$.
We propose that splittings satisfy the following conditions:
\begin{equation}\label{hierarchy-m4}
\begin{split}
&\tilde\epsilon_1\ll \tilde\epsilon_2\ll \tilde\epsilon_3,\\
&\frac{\tilde\epsilon_2^2}{\tilde\epsilon_1}\to 0,\quad\frac{\tilde\epsilon_3^2}{\tilde\epsilon_2}\to 0.\\
\end{split}
\end{equation}
Integration over $D_4$ can be rewritten as a double integral:
\begin{equation}
  \int_{D_4} dy dz =
  \int_{x+\tilde\epsilon_1}^{x+\tilde\epsilon_2-\tilde\epsilon_1}dy\int_{x+\tilde\epsilon_2}^{w-\tilde\epsilon_1}dz +
  \int_{x+\tilde\epsilon_2-\tilde\epsilon_1}^{w-\tilde\epsilon_2}dy\int_{y+\tilde\epsilon_1}^{w-\tilde\epsilon_1}dz.
\end{equation}
Doing some math this $m_4$-contribution can be estimated as
\begin{align}
  m_4^{\epsilon,E,E_1}(V_1,V_1,V_1,V_1) = O(\mathrm{splittings})+\ap O(\mathrm{splittings})+O(\ap^2),
\end{align}
where $O(\mathrm{splittings})$ is a term which vanishes as splittings go to zero according to \eqref{hierarchy-m4}.
Thus we conclude that due to lack of singularity in splittings at one contraction
and plenty of integrations the Yang-Mills equation is not affected by the $m_4$-contribution.
Similar reasoning is valid for $m_k$-contribution with $k\ge 5$ and
their compositions.
Thus we arrive to a conclusion that $m_k$-contribution with $k\ge4$ and their compositions 
do not affect the Yang-Mills equation 
at the leading order in $\ap$, although contribute to $\ap$-corrections.

\section{Conclusion}
In this paper we have constructed the maps that form $A_\infty$-algebra with a parameter.
By means of these maps it is possible to define the homotopical Maurer-Cartan equation,
find its symmetry and explore its properties.
Then using suitable parametrization we have reproduced the Yang-Mills equation
in a particular limit of the parameter.
Also we have identified $\ap$-correction linear in $\ap$ and
have suggested a calculation method for $\ap$-corrections.
Though calculations using this method becomes very complicated as we are trying to
find higher order $\ap$-correction,
we believe that there still exists a way that is able to take into account all corrections.
This approach with slight modification can be applied to the local operators inserted in the bulk of the worldsheet
thus providing an interpretation of string gravity equations in the same spirit.

Presented approach may be seen as a "string field theory" in sense that local operator $V$
may be considered as a conformal representation of a string field.
Thus expressions for $V_k$ may serve as appropriate basis for string field expansion.

\section{Acknowledgments}
We are grateful to A.S. Losev for motivation, inspiration and illuminating discussions during the work on the project.
Work of DG and PK was supported in part by 
the Russian Ministry of Science and Education (contract No 14.740.11.0710)
and 
grant RFBR 08-01-00798.

\begin{appendix}
\section{Verification of the quadratic relations of $A_\infty$-algebra} \label{section:quad-rel}
\begin{figure}[h!]
\begin{center}
  \begin{picture}(0,0)%
\includegraphics{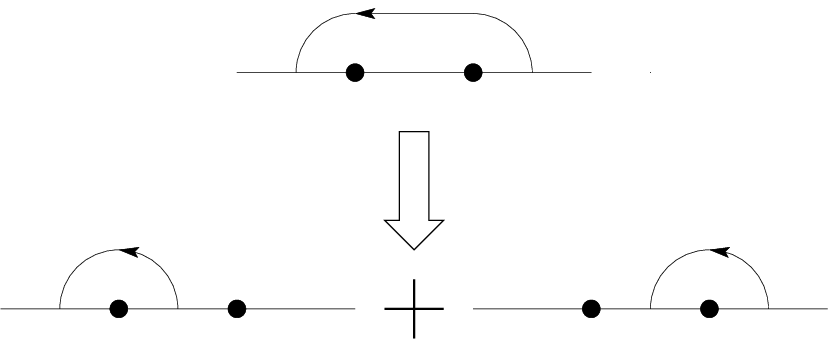}%
\end{picture}%
\setlength{\unitlength}{1243sp}%
\begingroup\makeatletter\ifx\SetFigFont\undefined%
\gdef\SetFigFont#1#2#3#4#5{%
  \reset@font\fontsize{#1}{#2pt}%
  \fontfamily{#3}\fontseries{#4}\fontshape{#5}%
  \selectfont}%
\fi\endgroup%
\begin{picture}(12624,5833)(-11,-5327)
\put(5131,-1546){\makebox(0,0)[lb]{\smash{{\SetFigFont{9}{10.8}{\rmdefault}{\mddefault}{\updefault}{\color[rgb]{0,0,0}$A$}%
}}}}
\put(6991,-1546){\makebox(0,0)[lb]{\smash{{\SetFigFont{9}{10.8}{\rmdefault}{\mddefault}{\updefault}{\color[rgb]{0,0,0}$B$}%
}}}}
\put(5311,119){\makebox(0,0)[lb]{\smash{{\SetFigFont{9}{10.8}{\rmdefault}{\mddefault}{\updefault}{\color[rgb]{0,0,0}$\int j_{BRST}$}%
}}}}
\put(3376,-5191){\makebox(0,0)[lb]{\smash{{\SetFigFont{9}{10.8}{\rmdefault}{\mddefault}{\updefault}{\color[rgb]{0,0,0}$B$}%
}}}}
\put(9796,-3346){\makebox(0,0)[lb]{\smash{{\SetFigFont{9}{10.8}{\rmdefault}{\mddefault}{\updefault}{\color[rgb]{0,0,0}$\int j_{BRST}$}%
}}}}
\put(8776,-5191){\makebox(0,0)[lb]{\smash{{\SetFigFont{9}{10.8}{\rmdefault}{\mddefault}{\updefault}{\color[rgb]{0,0,0}$A$}%
}}}}
\put(10576,-5191){\makebox(0,0)[lb]{\smash{{\SetFigFont{9}{10.8}{\rmdefault}{\mddefault}{\updefault}{\color[rgb]{0,0,0}$B$}%
}}}}
\put(1576,-5191){\makebox(0,0)[lb]{\smash{{\SetFigFont{9}{10.8}{\rmdefault}{\mddefault}{\updefault}{\color[rgb]{0,0,0}$A$}%
}}}}
\put(1576,-3331){\makebox(0,0)[lb]{\smash{{\SetFigFont{9}{10.8}{\rmdefault}{\mddefault}{\updefault}{\color[rgb]{0,0,0}$\int j_{BRST}$}%
}}}}
\end{picture}%
  \caption{The second relation.}
  \label{fig:Q-m2-uhp}
\end{center}
\end{figure}

In this section we check that the maps $m_k$ do form $A_\infty$-algebra by verifying the first three quadratic relations of $A_\infty$-algebra.
We note that this derivation can be easily adapted to proving quadratic relations for $M_k$ by drawing paths of integration on corresponding surface instead of the UHP.
This technique can be easily generalized to proving of all quadratic relations, not just three of them.
The first relation $Q^2=0$ is obvious. The second relation:
\begin{equation}
  Qm_2^\epsilon(A,B)=m_2^\epsilon(QA,B)+(-1)^Bm_2^\epsilon(A,QB).
\end{equation}
To prove it we can write $Q$ as a half-contour integral $Q=\int j_{BRST}$. Since $A$ and $B$ are inserted at distinct points we
can rewrite the integral as a sum of two integrals (See fig. \ref{fig:Q-m2-uhp}). The path of the first integral goes
around the point where $A$ is inserted, and the second path goes around point where $B$ is inserted.
One method to prove that we can move the path of integrations is doubling trick. It proves that we can move the path as long as
both ends of the path remain on real axis.

\begin{figure}[h!]
\begin{center}
  \begin{picture}(0,0)%
\includegraphics{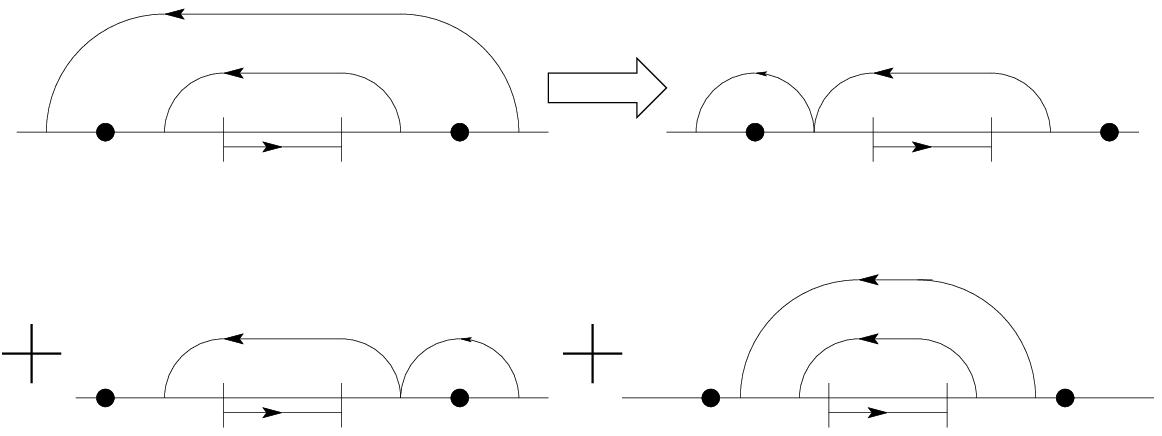}%
\end{picture}%
\setlength{\unitlength}{1243sp}%
\begingroup\makeatletter\ifx\SetFigFont\undefined%
\gdef\SetFigFont#1#2#3#4#5{%
  \reset@font\fontsize{#1}{#2pt}%
  \fontfamily{#3}\fontseries{#4}\fontshape{#5}%
  \selectfont}%
\fi\endgroup%
\begin{picture}(17595,7558)(-932,-5552)
\put(451,-1096){\makebox(0,0)[lb]{\smash{{\SetFigFont{9}{10.8}{\rmdefault}{\mddefault}{\updefault}{\color[rgb]{0,0,0}$A$}%
}}}}
\put(1576,1619){\makebox(0,0)[lb]{\smash{{\SetFigFont{9}{10.8}{\rmdefault}{\mddefault}{\updefault}{\color[rgb]{0,0,0}$\int j_{BRST}$}%
}}}}
\put(2476,614){\makebox(0,0)[lb]{\smash{{\SetFigFont{9}{10.8}{\rmdefault}{\mddefault}{\updefault}{\color[rgb]{0,0,0}$\int b$}%
}}}}
\put(5896,-1096){\makebox(0,0)[lb]{\smash{{\SetFigFont{9}{10.8}{\rmdefault}{\mddefault}{\updefault}{\color[rgb]{0,0,0}$C$}%
}}}}
\put(3016,-1366){\makebox(0,0)[lb]{\smash{{\SetFigFont{9}{10.8}{\rmdefault}{\mddefault}{\updefault}{\color[rgb]{0,0,0}$B$}%
}}}}
\put(10351,-1096){\makebox(0,0)[lb]{\smash{{\SetFigFont{9}{10.8}{\rmdefault}{\mddefault}{\updefault}{\color[rgb]{0,0,0}$A$}%
}}}}
\put(12376,614){\makebox(0,0)[lb]{\smash{{\SetFigFont{9}{10.8}{\rmdefault}{\mddefault}{\updefault}{\color[rgb]{0,0,0}$\int b$}%
}}}}
\put(15796,-1096){\makebox(0,0)[lb]{\smash{{\SetFigFont{9}{10.8}{\rmdefault}{\mddefault}{\updefault}{\color[rgb]{0,0,0}$C$}%
}}}}
\put(12916,-1366){\makebox(0,0)[lb]{\smash{{\SetFigFont{9}{10.8}{\rmdefault}{\mddefault}{\updefault}{\color[rgb]{0,0,0}$B$}%
}}}}
\put(9676,719){\makebox(0,0)[lb]{\smash{{\SetFigFont{9}{10.8}{\rmdefault}{\mddefault}{\updefault}{\color[rgb]{0,0,0}$\int j_{BRST}$}%
}}}}
\put(9676,-5146){\makebox(0,0)[lb]{\smash{{\SetFigFont{9}{10.8}{\rmdefault}{\mddefault}{\updefault}{\color[rgb]{0,0,0}$A$}%
}}}}
\put(15121,-5146){\makebox(0,0)[lb]{\smash{{\SetFigFont{9}{10.8}{\rmdefault}{\mddefault}{\updefault}{\color[rgb]{0,0,0}$C$}%
}}}}
\put(12241,-5416){\makebox(0,0)[lb]{\smash{{\SetFigFont{9}{10.8}{\rmdefault}{\mddefault}{\updefault}{\color[rgb]{0,0,0}$B$}%
}}}}
\put(12211,-3391){\makebox(0,0)[lb]{\smash{{\SetFigFont{9}{10.8}{\rmdefault}{\mddefault}{\updefault}{\color[rgb]{0,0,0}$\int b$}%
}}}}
\put(11926,-2431){\makebox(0,0)[lb]{\smash{{\SetFigFont{9}{10.8}{\rmdefault}{\mddefault}{\updefault}{\color[rgb]{0,0,0}$\int j_{BRST}$}%
}}}}
\put(451,-5146){\makebox(0,0)[lb]{\smash{{\SetFigFont{9}{10.8}{\rmdefault}{\mddefault}{\updefault}{\color[rgb]{0,0,0}$A$}%
}}}}
\put(2476,-3436){\makebox(0,0)[lb]{\smash{{\SetFigFont{9}{10.8}{\rmdefault}{\mddefault}{\updefault}{\color[rgb]{0,0,0}$\int b$}%
}}}}
\put(5896,-5146){\makebox(0,0)[lb]{\smash{{\SetFigFont{9}{10.8}{\rmdefault}{\mddefault}{\updefault}{\color[rgb]{0,0,0}$C$}%
}}}}
\put(3016,-5416){\makebox(0,0)[lb]{\smash{{\SetFigFont{9}{10.8}{\rmdefault}{\mddefault}{\updefault}{\color[rgb]{0,0,0}$B$}%
}}}}
\put(5176,-3331){\makebox(0,0)[lb]{\smash{{\SetFigFont{9}{10.8}{\rmdefault}{\mddefault}{\updefault}{\color[rgb]{0,0,0}$\int j_{BRST}$}%
}}}}
\end{picture}%
  \caption{The third relation.}
  \label{fig:Q-m3-uhp}
\end{center}
\end{figure}

The third relation is
\begin{equation}
\begin{split}
  &Q m_3^{\epsilon_1,\epsilon_2}(A,B,C)= -m_3^{\epsilon_1,\epsilon_2}(QA,B,C)-(-1)^Am_3^{\epsilon_1,\epsilon_2}(A,QB,C)-(-1)^{A+B}m_3^{\epsilon_1,\epsilon_2}(A,B,QC) \\
    &\quad\quad\quad + m_2^{\epsilon_2}(A,m_2^{\epsilon_1}(B,C))-m_2^{\epsilon_2}(m_2^{\epsilon_1}(A,B),C)
\end{split}
\end{equation}
We use the same techinque to rewrite $\int j_{BRST}$ into a sum of several integrals (see fig. \ref{fig:Q-m3-uhp}).
This figure illustrates the following derivation:
\begin{equation}
\begin{split}
  &Q m_3^{\epsilon_1,\epsilon_2}(A,B,C)=\\
  &\quad -m_3^{\epsilon_1,\epsilon_2}(QA,B,C)-(-1)^{A+B}m_3^{\epsilon_1,\epsilon_2}(A,B,QC)+A(x)Q\Bigl(\int dzb(z) \int dyB(y)\Bigr)C(u)=\\
  &\quad -m_3^{\epsilon_1,\epsilon_2}(QA,B,C)-(-1)^{A+B}m_3^{\epsilon_1,\epsilon_2}(A,B,QC)+A(x)\int dzT(z) \int dyB(y)C(u)\\
    &\quad\quad-(-1)^Am_3^{\epsilon_1,\epsilon_2}(A,QB,C)=\\
  &\quad -m_3^{\epsilon_1,\epsilon_2}(QA,B,C)-(-1)^Am_3^{\epsilon_1,\epsilon_2}(A,QB,C)-(-1)^{A+B}m_3^{\epsilon_1,\epsilon_2}(A,B,QC)\\
    &\quad\quad+A(x)\int dy\partial B(y)C(u)=\\
  &\quad -m_3^{\epsilon_1,\epsilon_2}(QA,B,C)-(-1)^Am_3^{\epsilon_1,\epsilon_2}(A,QB,C)-(-1)^{A+B}m_3^{\epsilon_1,\epsilon_2}(A,B,QC)\\
    &\quad\quad+m_2^{\epsilon_2}(A,m_2^{\epsilon_1}(B,C))-m_2^{\epsilon_2}(m_2^{\epsilon_1}(A,B),C)
\end{split}
\end{equation}
We note that this derivation is correct for any finite values of splittings, it doesn't require any hierarchy of the splittings.

\section{Note on redefinition}\label{section:redef}
Here we denote $\lambda=\frac\ap2\log\epsilon^2$.
Consider
\begin{equation}
\begin{split}
U&=f(X)g(X)+\lambda\partial_\mu f(X)\partial_\mu g(X)+\frac{\lambda^2}{2}\partial_\mu\partial_\nu f(X)\partial_\mu\partial_\nu g(X)+\ldots\\
&=\exp\Bigl(\lambda\frac{\partial}{\partial X^\mu}\frac{\partial}{\partial Y^\mu}\Bigr)\bigl(f(X)g(Y)\bigr)\Bigr|_{Y=X}
\end{split}
\end{equation}
where $f(X)$ and $g(X)$ are functions.
Using the following redefinition:
\begin{equation}
\begin{split}
f(X)&=\exp\Bigl(\frac{\lambda}{2}\frac{\partial}{\partial X^\mu}\frac{\partial}{\partial X^\mu}\Bigr)\tilde f(X)\\
g(X)&=\exp\Bigl(\frac{\lambda}{2}\frac{\partial}{\partial X^\mu}\frac{\partial}{\partial X^\mu}\Bigr)\tilde g(X)
\end{split}
\end{equation}
we can rewrite $U$ as
\begin{equation}
U=\exp\Bigl(\frac{\lambda}{2}\frac{\partial}{\partial X^\mu}\frac{\partial}{\partial X^\mu}\Bigr)\bigl(\tilde f(X)\tilde g(X)\bigr)
\end{equation}
This is also true for product of three and more functions:
\begin{equation}
\begin{split}
U&=f(X)g(X)h(X)+\lambda\partial_\mu f(X)\partial_\mu g(X)h(X)\\
&\quad+\lambda\partial_\mu f(X)g(X)\partial_\mu h(X)+\lambda f(X)\partial_\mu g(X)\partial_\mu h(X)+\ldots\\
&=\exp\Bigl(\lambda\frac{\partial}{\partial X^\mu}\frac{\partial}{\partial Y^\mu}\Bigr)
  \exp\Bigl(\lambda\frac{\partial}{\partial X^\mu}\frac{\partial}{\partial Z^\mu}\Bigr)
  \exp\Bigl(\lambda\frac{\partial}{\partial Y^\mu}\frac{\partial}{\partial Z^\mu}\Bigr)
  \bigl(f(X)g(Y)h(Z)\bigr)\Bigr|_{Z=Y=X}\\
&=\exp\Bigl(\frac{\lambda}{2}\frac{\partial}{\partial X^\mu}\frac{\partial}{\partial X^\mu}\Bigr)\bigl(\tilde f(X)\tilde g(X)\tilde h(X)\bigr)
\end{split}
\end{equation}
The redefinition $f\to\tilde f$ is similar to redefinitions made in \cite{YM-from-OSFT} when obtaining the Yang-Mills action from the
bosonic open string field theory.

\end{appendix}

\end{document}